\documentclass[letterpaper,journal]{IEEEtran} 
\usepackage{amsmath,amsfonts}
\usepackage{algorithmic}
\usepackage{algorithm}
\usepackage{array}
\usepackage{adjustbox}
\usepackage[caption=false,font=normalsize,labelfont=sf,textfont=sf]{subfig}

\usepackage{textcomp,enumitem}
\usepackage{stfloats}
\usepackage{url,booktabs}
\usepackage{verbatim}
\usepackage{graphicx}
\usepackage{cite,mathcomSTEv4}
 \usepackage{enumitem}
 \usepackage{graphicx}
 \usepackage{xspace}
 
\usepackage{xcolor}
\usepackage{amssymb}

\usepackage{xcolor}         
\definecolor{links}{rgb}{0.0,0,0.9}   
\definecolor{urls}{rgb}{0,0,0.9}    
\definecolor{cites}{rgb}{0.0,0.0,0.9}   
\usepackage[colorlinks,hyperindex,linkcolor=links,citecolor=cites,urlcolor=urls]{hyperref} 

\definecolor{rocpurple}{RGB}{128, 0, 128}

\definecolor{myDarkGreen}{RGB}{0,100,0} 
\definecolor{myDarkBlue}{RGB}{0,0,139} 
\definecolor{myDarkGray}{gray}{0.25} 

\newcommand{\acro}{\texttt{NARRAS}\xspace }

\begin{document}

\title{
\acro:
Edge-Triggered Distributed Inference for CSI-Based Localization in Vehicular IoT Networks
}
\author{Rodrigo Oliver, Ricardo Vazquez Alvarez, Alejandro Lancho, and Stefano Rini%
\thanks{R. Oliver, R. Vazquez, and A. Lancho are with the Signal Theory and Communications Department, Universidad Carlos III de Madrid, Leganés, 28911, Spain, and also with the Gregorio Marañón Health Research Institute, Madrid, 28007, Spain (e-mail: \{roliver, ricvazqu\}@pa.uc3m.es; alancho@ing.uc3m.es). S. Rini is with the Department of Electrical and Computer Engineering, National Yang-Ming Chiao-Tung University (NYCU), Hsinchu, Taiwan (e-mail: stefano.rini@nycu.edu.tw) and with the German Aerospace Center (DLR), Oberpfaffenhofen, Germany (e-mail stefano.rini@dlr.de). This work was supported in part by the Comunidad de Madrid under Grants 2023-T1/COM-29065 and SYG-2024/COM-870, and in part by the Ministerio de Ciencia, Innovación y Universidades, Spain, under Grant PID2023-148856OA-I00, funded by MICIU/AEI/10.13039/501100011033 and by ERDF/EU. 
It was also supported in part by the NSTC grant 114-2923-E-A49 -006 -MY3 and 113-2923-E-A49 -001 -MY2 and by MARC, the MediaTEK Advanced Research Center, with grant number 114A540531.
}%
}

\maketitle

\begin{abstract}
CSI-based localization with spatially distributed antenna arrays exposes a basic resource trade-off. Each array can provide a rich view of the channel, but forwarding observations from all arrays to a fusion center is wasteful when only a few carry useful information, and the shared uplink supports only a limited number of simultaneous transmissions. We let each array decide locally whether its current observation is worth reporting, subject to a budget on the average number of active transmitters. We refer to this abstraction as \emph{Edge-Triggered Distributed Inference} (ETDI). It captures a broader class of task-oriented communication problems where resource-constrained devices share an access channel for a common inference task.
We instantiate ETDI for CSI-based localization, a common scenario in vehicular IoT networks. Spatially distributed remote antenna arrays (RAAs) encode local channel state information (CSI) from user equipment (UE) transmissions into latent features, and the fusion center estimates the UE position from the subset of reported features. We propose \acro, a decentralized reporting policy in which each RAA combines a recurrent summary of its recent observations with a memory of the last latent it transmitted. Training controls an explicit activity budget through differentiable activity penalties and validation-calibrated deterministic thresholds, and uses channel-chart regularization to shape the latent geometry.
Experiments show that, at comparable uplink activity, \acro improves localization accuracy over learned and heuristic sparse-reporting strategies, while dense full-report models remain useful budget-free references. In low-activity regimes, chart regularization further reduces high-percentile localization errors, suggesting that geometry-aware latent representations are more robust under sparse reporting.
\end{abstract}


\begin{IEEEkeywords}
Edge AI, Distributed inference, Resource-constrained IoT, Grant-free random access, Channel charting.
\end{IEEEkeywords}


\section{Introduction}
\label{sec:introduction}

\IEEEPARstart{T}{he} growth of vehicular and low-altitude IoT networks is pushing intelligence from centralized cloud processors toward distributed edge nodes, such as onboard units, roadside sensors, UAVs, and remote radio heads. These nodes must support several real-time tasks under tight communication, computation, and energy constraints: perception, coordination, localization, and access management.
In many practical deployments, however, not all local observations are equally informative. Only a subset of devices may observe measurements that are highly relevant for the task, while others provide redundant or low-utility information. Continuously polling all devices can therefore waste communication resources and increase contention over the shared medium. 
This motivates architectures in which devices decide locally whether their information is worth reporting.
When local reporting decisions are part of the system design, contention over the shared uplink becomes a central bottleneck. 
A natural abstraction for this effect is a random access channel (RACH) with limited throughput, modeled here through a constraint on the average number of active transmitters. This leads to the system model studied in this work, which we refer to as \emph{Edge-Triggered Distributed Inference} (ETDI). In ETDI, edge devices execute local reporting policies that decide whether to access the shared uplink under an explicit communication budget.

To ground ETDI in a practical radio-localization setting relevant to vehicular and industrial IoT, we consider channel state information (CSI)-based indoor localization using the DICHASUS dataset \cite{Euchner2021-mx,Euchner2022-am,Euchner2023-wu}.
Indoor localization is important for applications ranging from navigation and asset tracking to autonomous systems and network management. While specialized systems such as ultra-wideband (UWB) anchors or fingerprinting campaigns can achieve high accuracy, they often require significant signaling and calibration overhead \cite{Djosic2021-hl}. By contrast, modern wireless systems already estimate CSI during normal uplink operation \cite{Larsson2014-bf,Liu2018-qh}, and CSI captures rich multipath effects induced by the environment \cite{Wu2013-zi,Studer2018-dk}.
This application is particularly well suited for ETDI. In distributed antenna deployments, such as roadside units and distributed sensing infrastructure supporting V2X services, CSI observations across remote antenna arrays (RAAs) often have very different relevance for localization. Arrays with favorable line-of-sight conditions or rich multipath interactions may provide highly informative representations, while others observe weak or redundant channel realizations. Rather than forwarding full-resolution CSI from all RAAs, each RAA can extract a compact task-oriented representation and decide autonomously whether to report it. The fusion center then estimates the user equipment (UE) location from the subset of received reports. This setting naturally exposes the trade-off between localization accuracy and communication cost under a shared uplink activity constraint. Although our experiments use measured indoor distributed CSI, CSI localization here serves as a representative edge inference task, and the same design principles apply to vehicular and low-altitude IoT deployments in which ground or aerial nodes must support radio-based localization or access management with limited communication, computation, and energy resources.

\subsection{Related Work}
\label{sec:related_work}

The ETDI framework lies at the intersection of CSI-based localization, distributed antenna processing, and communication-constrained uplink access.

\paragraph{CSI-based localization and channel charting}
Supervised fingerprinting learns direct mappings from CSI to absolute position using labeled data \cite{Foliadis2024-yn}. Channel charting, introduced in \cite{Studer2018-dk,Medjkouh2018-sh}, instead learns low-dimensional representations that preserve local spatial neighborhoods in a self-supervised manner, with subsequent works improving chart quality through triplet-based, deep-manifold, and autoencoder objectives \cite{Ferrand2020-of,Lei2019-qr}. More recent studies refine CSI dissimilarity metrics in distributed massive multiple-input multiple-output (MIMO) settings and extend charting to channel prediction \cite{Stephan2024-tw}, align charts with physical coordinates using anchors or digital twins \cite{Taner2025-jo}, or embed chart representations directly into physical space \cite{Lei2019-qr,Palhares2025-wg}.

\paragraph{Centralized processing and fronthaul constraints}
Most CSI-based learning methods process CSI from distributed antennas jointly at a central node \cite{Studer2018-dk,Miretti2024-ge,Mobini2026-uo}, implicitly assuming abundant fronthaul capacity and continuous reporting. In practice, distributed antenna systems are often limited by fronthaul and reporting overhead \cite{Garcia-Saavedra2018-fn,Wiffen2021-cn}, which motivates selective reporting and compressed representations.

\paragraph{Grant-free access and reporting reduction}
Grant-free uplink random access is central to massive machine-type communication, with a substantial literature on activity detection, collision mitigation, and channel estimation under shared access \cite{Liu2018-qh}. In distributed antenna systems, reporting load has been reduced using signal-to-noise-ratio (SNR)- or geometry-based selection, as well as learning-based remote radio head selection \cite{Salihu2022-hk}. Channel charting has also been used to guide resource allocation by exploiting the geometry learned from CSI \cite{Ribeiro2020-gy,Che2024-ma}, and reconstruction-oriented autoencoders have been studied for efficient CSI feedback, particularly in frequency-division duplex (FDD) scenarios \cite{Guo2022-ih}.

What remains less explored is a unified design that learns task-oriented representations for localization, controls explicit uplink activity budgets inspired by grant-free access, and regularizes the communicated representation using channel charting principles within a decentralized reporting architecture. The ETDI framework addresses this gap by jointly learning local encoders, reporting policies, and a permutation-invariant fusion model under an explicit average uplink constraint.

\subsection{Contributions}
\label{sec:contributions}

Using measured distributed CSI from DICHASUS industrial scenarios, this paper makes the following contributions. First, we propose \acro, a decentralized reporting mechanism in which each RAA decides locally whether to report based on a learned probability computed from its current latent representation, its local temporal state, and the last communicated latent. Second, we train a dense full-report localization model using supervised localization and CSI-neighborhood channel-chart regularization, then use it as both a warm start and a frozen teacher for sparse ETDI models; sparse training combines localization, channel-chart regularization, and dense-teacher distillation. Third, we evaluate \acro under explicit average-activity budgets and compare it against a set of internal reference policies defined within the same architecture and protocol: dense reporting, random reporting, received-power gating, latent-drift gating, and principal-component-analysis (PCA)-based drift gating, together with simpler ETDI triggers. All policies share a common data split, calibration protocol, and fusion interface, and the evaluation reports localization error, realized activity, geometry preservation, deployment overhead, and spatial error structure. Together, these contributions show that learning \emph{when} to report---not only \emph{what} to encode---yields a markedly better localization--communication trade-off than heuristic and non-learned sparse reporting, approaching dense full-report accuracy while keeping the shared uplink lightly loaded, the regime that matters under grant-free access in vehicular IoT.

\subsection{Paper Organization}
\label{sec:paper_organization}

The remainder of the paper is organized as follows. Section~\ref{sec:sys_model} introduces the ETDI system model and notation; Section~\ref{sec:cc_problem} specializes it to CSI-based indoor localization. Section~\ref{sec:proposed_solution} presents \acro. Section~\ref{sec:exp_setup} describes the experimental protocol, compared methods, and metrics, and Section~\ref{sec:results} reports the results. Section~\ref{sec:conclusion} concludes.

\paragraph*{Notation}
Calligraphic letters denote sets. For a positive integer \(P\), let \([P] \triangleq \{1, \ldots, P\}\). Device indices use \(r \in [R]\) and sample indices use \(n \in [N]\), so subscripts index devices and superscripts \((n)\) index samples. Bold lowercase letters denote vectors and bold uppercase letters denote matrices, and \([\mathbf{a}\,;\,\mathbf{b}]\) denotes concatenation. For variables indexed by devices \(r \in [R]\), we write \(x_{1:R} = (x_1,\ldots,x_R)\).
The expectation is denoted by \(\mathbb{E}[\cdot]\), \(\mathbb{I}\{\cdot\}\) is the indicator function, \(|\mathcal{S}|\) is the cardinality of a set \(\mathcal{S}\), and \(\|\cdot\|_2\) and \(\|\cdot\|_F\) denote the Euclidean and Frobenius norms. For complex matrices, \((\cdot)^{\mathrm H}\) denotes Hermitian transpose and \(\mathrm{tr}(\cdot)\) the trace.

\begin{table}[t]
\centering
\caption{Notation summary for the ETDI system model in Section~\ref{sec:sys_model}.}
\label{tab:notation_etdi}
\begin{tabular}{ll ll}
\toprule
 & Meaning &  & Meaning \\
\midrule
$R$ & Num. of RAAs/devices
& $N$ & Num.  of samples \\

$L$ & Observation dim.
& $d$ & Latent dim. \\

$T$ & Prediction dim. 
& $K$ & Activity budget \\

$\mathbf{H}_r$ & Observation (real-stacked CSI)
& $\mathbf{z}_r$ & Latent representation \\

$\mu_r$ & Reporting policy
& $\pi_r$ & Report probability \\

$c_r$ & Reporting decision
& $\mathcal{Z}$ & Source-indexed report set \\

$f_r$ & Local encoder
& $g$ & Fusion model \\

$\mathbf{y}$ & Ground-truth label 
& $\hat{\mathbf{y}}$ & Predicted global output \\

$\mathcal{A}$ & Active device set
& $M$ & Num. active devices \\
\bottomrule
\end{tabular}
\end{table}
\section{Edge-Triggered Distributed Inference (ETDI) Framework}
\label{sec:sys_model}

\begin{figure}[t]
    \centering
    \includegraphics[width=0.88\columnwidth]{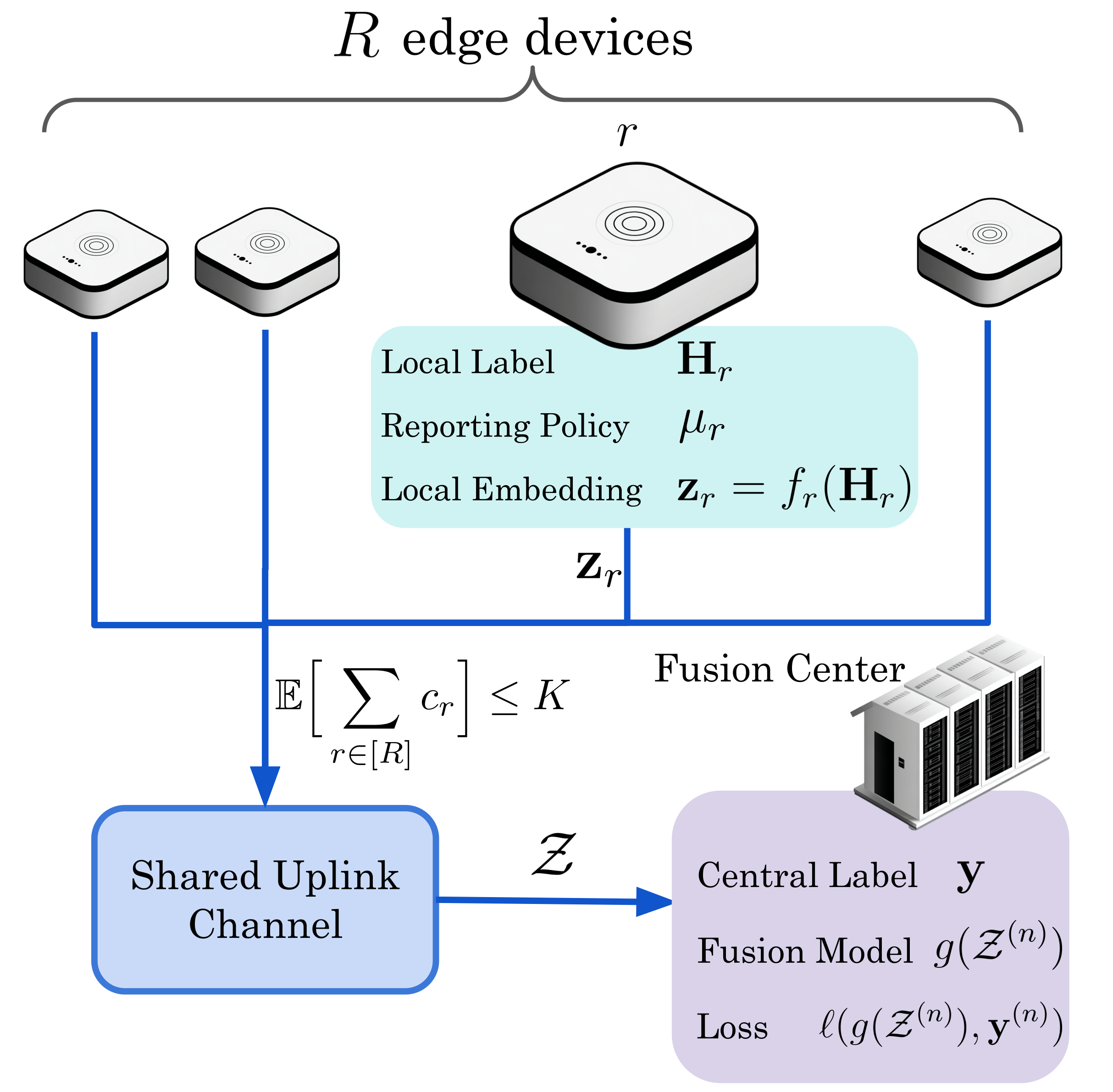}
    \vspace{-0.25cm}
    \caption{
        Edge-triggered distributed inference (ETDI) over a shared uplink. Each device \(r \in [R]\) observes \(\mathbf{H}_r\), computes \(\mathbf{z}_r=f_r(\mathbf{H}_r)\), and makes a binary reporting decision \(c_r\). The uplink activity target is \(\mathbb{E}[\sum_{r \in [R]} c_r] \leq K\), so the fusion center receives the source-indexed subset \(\mathcal{Z} = \{(r,\mathbf{z}_r) : c_r=1,\ r\in[R]\}\) and produces the estimate \(\hat{\mathbf{y}} = g(\mathcal{Z})\).
    }
    \label{fig:etdi_overview}
\end{figure}

We consider a distributed inference setting in which \(R\) edge devices observe local data and communicate with a central fusion center over a shared uplink channel. Each device decides autonomously whether to transmit based only on its local observation, and reporting takes place over a contention-based uplink that we model as a grant-free RACH with an average throughput constraint. The expected number of active transmitting devices per inference is thus bounded, which captures the main trade-off in ETDI: reducing communication while maintaining inference performance. The rest of this section formalizes the four components of the model, i.e., local observations, encoding and reporting, the shared uplink constraint, and fusion, which are illustrated in Fig.~\ref{fig:etdi_overview} and summarized in Table~\ref{tab:notation_etdi}.

\paragraph{Distributed Observations}
Each device \(r \in [R]\) observes a local measurement $\mathbf{H}_r \in \mathbb{R}^{L}$, which contains task-relevant information available only at that device. The fusion center aims to infer a global variable $\mathbf{y} \in \mathbb{R}^{T}$, which represents the target of the inference task.

For learning, we assume access to a dataset
\begin{equation}
\label{eq:etdi_labeled_data}
\mathcal{D}
=
\left\{
(\{\mathbf{H}_r^{(n)}\}_{r \in [R]}, \mathbf{y}^{(n)})
\right\}_{n \in [N]},
\end{equation}
where $\mathbf{H}_r^{(n)}$ denotes the observation at device $r$ for sample $n$, and $\mathbf{y}^{(n)}$ denotes the associated central label.

\paragraph{Local Encoding and Reporting}
Each device processes its observation locally before deciding whether to transmit. In particular, device $r$ computes a latent representation
\begin{equation}
\label{eq:local_encoder}
\mathbf{z}_r = f_r(\mathbf{H}_r) \in \mathbb{R}^{d},
\end{equation}
where $f_r : \mathbb{R}^{L} \to \mathbb{R}^{d}$ is a local encoder.
The device then applies a local reporting policy, denoted by \(\mu_r\), using only information available at that device. In the implemented models this policy is parameterized by a report probability \(\pi_r \in (0,1)\), from which a binary communication decision $c_r \in \{0,1\}$ is produced, where \(c_r = 1\) means that device \(r\) reports its latent representation to the fusion center and \(c_r = 0\) means that it remains silent.
The source-indexed set of reports received at the fusion center is therefore
\begin{equation}
    \label{eq:received_latents_fc}
    \mathcal{Z}=\left\{
    (r,\mathbf{z}_r) : c_r=1,\; r\in[R]
    \right\}.
\end{equation}

\paragraph{Shared Uplink Constraint}
The devices communicate their latent reports to the fusion center over a shared uplink channel that is contention-based and unscheduled. To capture the effect of limited communication resources in a tractable way, we constrain the expected number of reporting devices:
\begin{equation}
\label{eq:average_throughput_constraint}
\mathbb{E}\left[\sum_{r \in [R]} c_r\right] \leq K,
\end{equation}
where $K \in \{1,\ldots,R\}$ is the communication budget. Smaller values of $K$ enforce sparser communication and require more selective reporting decisions.

\paragraph{Fusion and Inference}

The fusion center aggregates the received reports and performs inference on the global task. Since different devices may be active for different samples, the fusion input is a variable-cardinality unordered set. The fusion function can be viewed as a map accepting any subset of the \(R\) possible source-indexed latent reports,
\begin{equation}
\label{eq:fusion_map}
g:\left\{
\mathcal{S}\subseteq [R]\times\mathbb{R}^{d}:\ |\mathcal{S}|\leq R
\right\}
\rightarrow\mathbb{R}^{T}.
\end{equation}

For compact notation, we write \(g(\mathcal{Z})\). The resulting estimate is
\begin{equation}
\label{eq:fusion_estimate}
\hat{\mathbf{y}} = g(\mathcal{Z}).
\end{equation}

Permutation invariance is important because the set of active devices may change from one sample to another, and the fusion rule should not depend on the order in which the reports are received.

\paragraph{Modeling Assumptions}
The formulation above adopts several simplifying assumptions that isolate the core structure of ETDI: latent representations share a common dimension \(d\); the communication constraint is expressed only as an expected number of reporting devices, without per-device rates, bit budgets, or protocol overhead; the activity constraint is imposed in expectation rather than per-sample, reflecting the stochastic nature of contention over a RACH; reporting policies are fully decentralized, with no inter-device coordination; and the RACH is abstracted away from physical-layer effects such as collisions, modulation, capture, and protocol details. 

\section{CSI Localization and Channel Charting}
\label{sec:cc_problem}
\begin{figure}
    \centering
    {\includegraphics[width=\columnwidth]{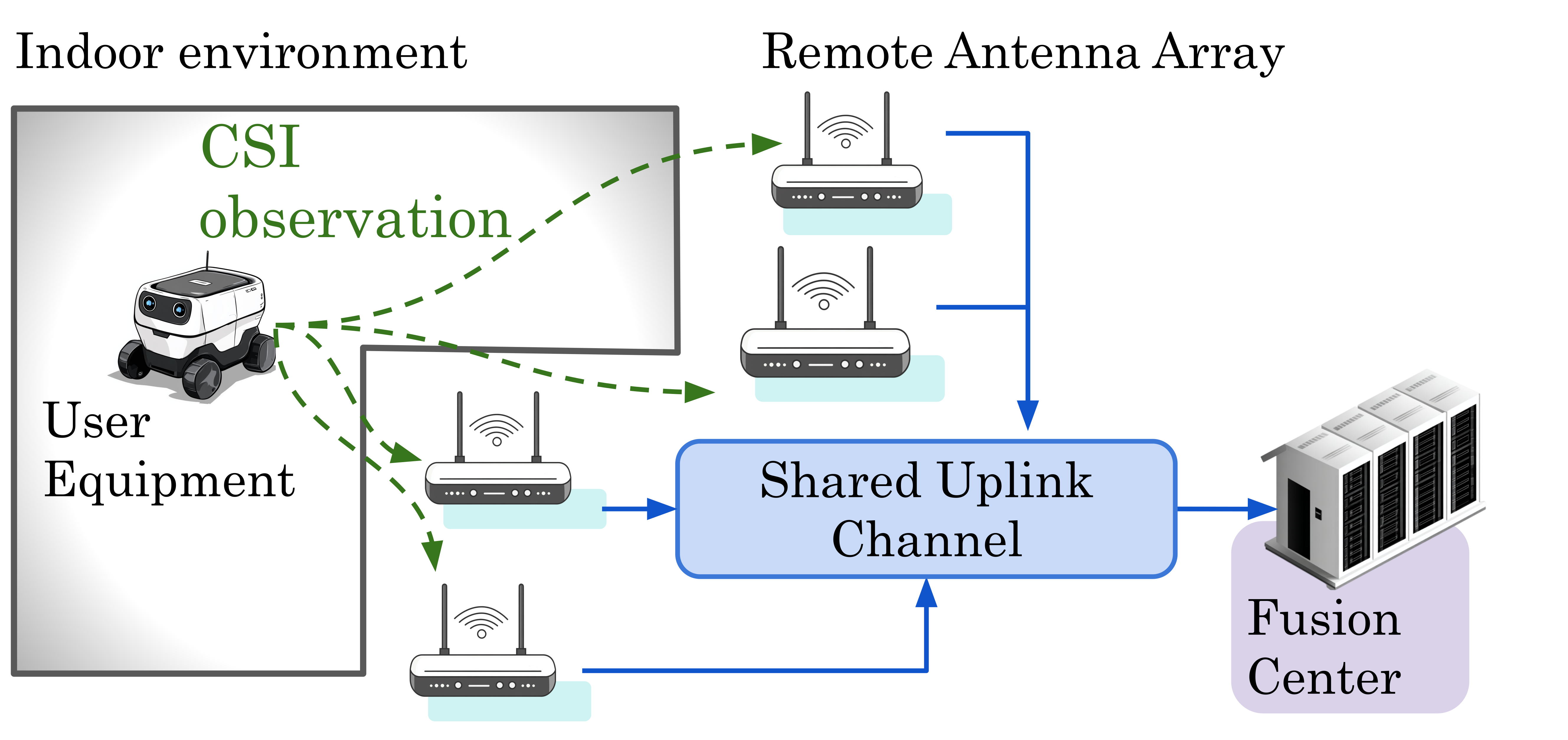}}
    \vspace{-0.5cm}
    \caption{
Distributed CSI localization with channel-chart regularization. A user equipment (UE) operating in an indoor environment induces local CSI observations at multiple RAAs (dashed green links). Active RAAs forward reports over a shared uplink channel (solid blue) to a fusion center, which aggregates the received information for UE localization and auxiliary channel-chart embedding.
    }
    \label{fig:channel_charting}
\end{figure}

We now specialize the ETDI framework of Section~\ref{sec:sys_model} to indoor localization from CSI collected by a network of spatially distributed antennas, as illustrated in Fig.~\ref{fig:channel_charting}. A single UE transmits, and each of the $R$ RAAs observes a local complex CSI snapshot. For compactness, we write $\mathbf{H}_r^{(n)} \in \mathbb{R}^{L}$ for the real-valued model input obtained by stacking the real and imaginary components of this snapshot. The target $\mathbf{y}^{(n)} \in \mathbb{R}^{T}$ is the UE position, with $T = 2$ in the experiments. Each RAA encodes its observation as $\mathbf{z}_r^{(n)} = f_r(\mathbf{H}_r^{(n)}) \in \mathbb{R}^d$, applies a local reporting policy $\mu_r$ with output $c_r^{(n)} \in \{0,1\}$, and contributes to the active set $\mathcal{A}^{(n)} = \{r : c_r^{(n)} = 1\}$. The fusion center then estimates $\hat{\mathbf{y}}^{(n)} = g(\mathcal{Z}^{(n)})$, where $\mathcal{Z}^{(n)} = \{(r,\mathbf{z}_r^{(n)}) : r \in \mathcal{A}^{(n)}\}$ and $g$ is permutation-invariant over source-indexed reports.

CSI varies smoothly with UE position, because local motion deforms the multipath structure in a continuous way: two CSI snapshots taken a short distance apart tend to resemble each other much more than two snapshots from distant points. The collection of CSI observations therefore tends to lie on a low-dimensional manifold whose neighborhood relations reflect physical proximity. Channel charting \cite{Studer2018-dk} turns this observation into a learning principle: it seeks a representation of the channel in which CSI samples that are similar (and therefore likely close in space) are also close in the representation. Crucially, this objective only needs CSI, not UE position. Pairs of ``should-be-close'' samples can be obtained directly from unlabeled CSI sequences, for instance from snapshots taken in quick succession by a moving UE. Charting can therefore be learned from abundant unlabeled data, while a smaller set of labeled samples is enough to tie the learned representation to physical coordinates.

We exploit this through an auxiliary map
\begin{equation}
\label{eq:chart_map}
g_c:
\left\{
\mathcal{S}\subseteq [R]\times\mathbb{R}^{d}:\ |\mathcal{S}|\leq R
\right\}
\rightarrow \mathbb{R}^{T_c},
\quad
\mathbf{y}_c^{(n)} = g_c(\mathcal{Z}^{(n)}),
\end{equation}
which produces a $T_c$-dimensional embedding (\(T_c = 16\) in the experiments) that preserves local CSI neighborhoods. The chart is not an additional label, but a representation learned from the CSI itself; it regularizes training and lets us check how well spatial geometry survives sparse reporting. The last point matters here because the fusion center only sees the subset $\mathcal{Z}^{(n)}$, which the RAAs choose locally and which varies from one sample to the next.

This logic fits the realities of practical measurement campaigns, in which only a fraction of the collected CSI can be paired with ground-truth UE positions. We use the DICHASUS distributed-MIMO CSI measurements \cite{Euchner2021-mx,Euchner2022-am,Euchner2023-wu}, in which a UE moves through an industrial environment while several spatially distributed RAAs record synchronized CSI. The dataset splits into
$\mathcal{D}_L = \{(\{\mathbf{H}_r^{(i)}\}_{r\in[R]}, \mathbf{y}^{(i)})\}_{i \in [N_L]}$,
containing the $N_L$ samples with known UE positions, and
$\mathcal{D}_U = \{\{\mathbf{H}_r^{(j)}\}_{r\in[R]}\}_{j \in [N_U]}$,
containing the $N_U$ unlabeled samples. Labeled samples anchor the chart to physical coordinates through the localization loss. In dense training, unlabeled samples enter through the channel-charting objective, which only requires CSI. In sparse training, unlabeled samples also contribute to the dense teacher's pooled and chart distillation terms, which require no position labels. Entire trajectories, rather than individual samples, are assigned to the labeled or unlabeled set; Section~\ref{sec:exp_setup} gives the full details.

The full learning problem then seeks local encoders, reporting policies, and a fusion map that minimize a task loss $\ell(\cdot,\cdot)$ on labeled samples while respecting the average uplink activity:
\begin{equation}
\begin{aligned}
\min_{f_{1:R},\,\mu_{1:R},\,g}
\quad &
\frac{1}{N_L}\sum_{n \in \mathcal{N}_L}
\mathbb{E}\!\left[
\ell\!\left(g(\mathcal{Z}^{(n)}), \mathbf{y}^{(n)}\right)
\right] \\
\text{s.t.} \quad &
\mathbb{E}\!\left[\sum_{r=1}^{R} c_r\right] \leq K,
\end{aligned}
\label{eq:etdi_learning_problem}
\end{equation}
where $\mathcal{N}_L$ is the set of labeled samples and the expectation is over the (possibly stochastic) reporting policies. Three concerns shape this problem at once. Geometry preservation comes from the channel-charting objective evaluated on $\mathcal{D}_L \cup \mathcal{D}_U$. Localization accuracy comes from $\ell$ on $\mathcal{D}_L$. Robustness to sparse reporting matters because the active set $\mathcal{A}^{(n)}$ changes across samples and shrinks as $K$ decreases.

Equation~\eqref{eq:etdi_learning_problem} states the operating constraint we want but it does not prescribe how to enforce it. In our implementation, we train each sparse model for a target budget $K$ using differentiable soft penalties on expected reporting activity and on all-silent events, followed by validation-based threshold calibration. Section~\ref{sec:proposed_solution} introduces the \acro architecture built around this problem. Simpler policy variants used for ablation are deferred to Section~\ref{sec:compared_methods}.

\section{Proposed Approach: \acro}
\label{sec:proposed_solution}

\begin{figure*}[t]
	\centering
 	\includegraphics[width=0.99\textwidth]{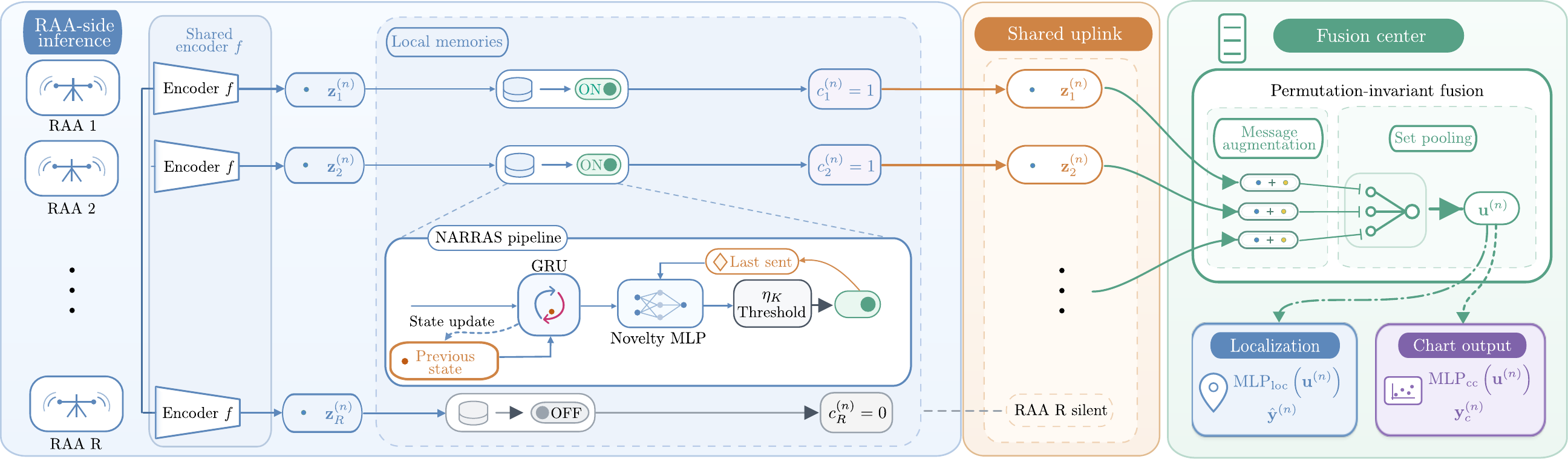}
	\caption{
    Architecture of \acro for activity-constrained CSI localization. Each RAA applies the shared encoder \(f\) to its local CSI observation and executes a decentralized recurrent novelty trigger using only local memories: a recurrent state summarizing recent observations and a last-sent latent reference. The trigger produces a report decision \(c_r^{(n)}\), so only active RAAs transmit their latent representations over the shared uplink. The fusion center receives a variable-cardinality set of reports, augments each report with source information, applies source-indexed attention pooling, and produces both the localization estimate \(\hat{\mathbf{y}}^{(n)}\) and the auxiliary channel-chart embedding \(\mathbf{y}_c^{(n)}\).
	}
	\label{fig:narras_architecture}
\end{figure*}

Building on the ETDI formulation in Sections~\ref{sec:sys_model} and~\ref{sec:cc_problem}, we now introduce a concrete architecture for CSI-based localization under three coupled requirements: decentralized device-side reporting, permutation-invariant fusion of a variable-cardinality set of reports, and explicit control of the average uplink activity. Fig.~\ref{fig:narras_architecture} gives an end-to-end view of the resulting method, \acro, which has four components: a shared edge encoder, a decentralized recurrent novelty trigger, a permutation-invariant fusion network, and a dense-to-sparse training procedure.

\paragraph{Shared Edge Encoder}
The encoder \(f_r\) introduced in Section~\ref{sec:cc_problem} is implemented as a lightweight convolutional network shared across all RAAs, i.e., \(f_r \equiv f\). Weight sharing reduces edge-side storage, simplifies deployment, and encourages latent representations from different RAAs to lie in a common feature space. Device-specific information is reintroduced later at the fusion stage through learned RAA embeddings.

\paragraph{Decentralized Reporting Policy}
The key design question in \acro is when an RAA should report its current latent representation. Our guiding principle is that a device should transmit only when its current observation is sufficiently novel relative to both its recent local history and what it has already communicated. Denote by \(\pi_r^{(n)} \in (0,1)\) the report probability and by \(c_r^{(n)} \in \{0,1\}\) the realized report/silence action for device \(r\) on sample \(n\). Throughout the trigger definitions, \(\sigma(x)=(1+\exp(-x))^{-1}\) denotes the logistic sigmoid.
During training, the hard communication action is implemented through a straight-through estimator,
\begin{equation}
\label{eq:ste_gate}
    c_r^{(n)}
    =
    \mathbb{I}
    \left\{
    \pi_r^{(n)} > v_r^{(n)}
    \right\},
    \quad
    v_r^{(n)} \sim \mathrm{Uniform}(0,1),
\end{equation}
so that \(c_r^{(n)}\) is Bernoulli with parameter \(\pi_r^{(n)}\). In backpropagation, the derivative of the hard gate is replaced by the derivative of the corresponding sigmoid relaxation.
At inference, communication is made deterministic through a single calibrated threshold \(\eta_K \in [0,1]\) associated with the target budget \(K\).
For each trained model, \(\eta_K\) is calibrated on validation using the same threshold for all RAAs and all samples. This threshold is not learned by gradient descent. The learned quantities are the report probabilities \(\pi_r^{(n)}\). For stateful policies, each device's memory evolves along a trajectory and resets at the start of the next.

\noindent
$\bullet$ \underline{Recurrent Novelty Trigger (RNT):}
In the main \acro configuration, the report probability is produced by the RNT. This trigger maintains two complementary local memories: a recurrent state \(\mathbf{o}_r^{(n)}\), updated by a gated recurrent unit (GRU) and summarizing recent observations, and a communication memory \(\mathbf{j}_r^{(n)}\) storing the last latent representation actually transmitted by device \(r\). A report is favored when the current latent deviates from the communicated reference while remaining interpretable in light of the recent local trajectory. The decision is then produced by a small multilayer perceptron (MLP). The local memory update and last-sent comparison are illustrated by the \acro pipeline inset in Fig.~\ref{fig:narras_architecture}.
These states evolve as
\begin{align}
\label{eq:rnt_states}
    \mathbf{o}_r^{(n)}
    & =
    \mathrm{GRU}_{\mathrm{pol}}
    \left(
    \mathbf{o}_r^{(n-1)}, \mathbf{z}_r^{(n)}
    \right),
    \quad
    \mathbf{o}_r^{(0)}=\mathbf{0},\\
    \mathbf{j}_r^{(n)}
    & =
    c_r^{(n)} \mathbf{z}_r^{(n)}
    +
    \left(1-c_r^{(n)}\right)\mathbf{j}_r^{(n-1)},
    \quad
    \mathbf{j}_r^{(0)}=\mathbf{0}.
\end{align}
Finally, the report probability is obtained as
\begin{equation}
\label{eq:rnt_prob}
    \pi_r^{(n)}
    =
    \sigma
    \left(
    \mathrm{MLP}_{\mathrm{rnn}}
    \left(
    \left[
    \mathbf{o}_r^{(n)}
    \,;\,
    \left\|
    \mathbf{z}_r^{(n)} - \mathbf{j}_r^{(n-1)}
    \right\|_2
    \right]
    \right)
    \right).
\end{equation}

Here, \(\mathbf{o}_r^{(n)}\) summarizes recent local evolution, whereas \(\mathbf{j}_r^{(n-1)}\) identifies what the fusion center most recently received from device \(r\). This combination lets the policy react to temporal novelty rather than to instantaneous magnitude alone.

\paragraph{Permutation-invariant Memoryless Fusion}
Recall from Section~\ref{sec:sys_model} that $g$ must be permutation-invariant. Here, the source index of each received report is used through a learned RAA embedding. The fusion-side processing shown in Fig.~\ref{fig:narras_architecture} therefore separates message augmentation from permutation-invariant set pooling.

Each reported latent is first augmented with a learned embedding \(\mathbf{a}_r\) associated with device \(r\):
\begin{equation}
\label{eq:msg_augment}
    \mathbf{m}_r^{(n)}
    =
    \left[
    \mathbf{z}_r^{(n)}
    \,;\,
    \mathbf{a}_r
    \right].
\end{equation}
These augmented reports are then aggregated through source-indexed attention pooling. In the implementation, the fusion module is evaluated on the fixed RAA axis: an attention logit and a set-element representation are computed for each RAA, the attention weights are masked by the report decisions, and the remaining weights are renormalized over the active RAAs. This is equivalent to permutation-invariant pooling over the set of source-indexed pairs \((r,\mathbf{z}_r)\) when the source index is preserved.
If no RAA reports, the fusion center uses a learned default representation $\mathbf{u}_{\varnothing}$.
Both the localization map \(g\) and the charting map \(g_c\) are realized from the same pooled representation.
This shared-backbone design is central to the method: charting acts as a structural regularizer of the same fused representation from which localization is predicted.

\paragraph{Dense-to-Sparse Training with Activity Regularization}
Training proceeds in two stages. First, we train a dense full-report model in which all RAAs are active for every sample. This dense model uses the same shared encoder, set fusion module, localization head, and chart head, but no sparse reporting trigger. Second, for each target activity budget \(K\), we train a sparse ETDI student initialized from the dense model. The dense model is then frozen and used as a teacher during sparse training.

Let \(\mathcal{B}\) denote a mini-batch, let \(\mathcal{V}_{\mathcal B}\) denote its valid timesteps, and let \(\mathcal{V}_{\mathcal B,L}\subseteq\mathcal{V}_{\mathcal B}\) denote the valid labeled timesteps. The supervised localization term is
\begin{equation}
    \label{eq:loc_batch_loss}
    \mathcal{L}_{\mathrm{loc}}
    =
    \frac{1}{\left|\mathcal{V}_{\mathcal B,L}\right|}
    \sum_{n \in \mathcal{V}_{\mathcal B,L}}
    \left\|
    \hat{\mathbf{y}}^{(n)}
    -
    \mathbf{y}^{(n)}
    \right\|_2^2.
\end{equation}

When \(|\mathcal{V}_{\mathcal B,L}|=0\), we set \(\mathcal{L}_{\mathrm{loc}}=0\).

To stabilize learning under sparse reporting, we regularize the same fused  representation with a channel-charting loss that encourages nearby CSI samples to remain neighbors in the chart space. This term is not a separate prediction task; rather, it preserves local geometry in the representation used for localization.
For each snapshot \(i\) and RAA \(r\), let \(\mathbf{X}_{i,r}\in\mathbb{C}^{A\times F}\) denote the antenna-by-subcarrier CSI matrix after flattening the local antenna grid, where \(A\) is the number of local antenna elements and \(F\) is the number of subcarriers. The normalized spatial covariance descriptor is
\begin{equation}
	\label{eq:cov_descriptor}
	\mathbf{C}_{i,r}
	=
	\frac{
	\mathbf{X}_{i,r}\mathbf{X}_{i,r}^{\mathrm H}
	}{
	\|\mathbf{X}_{i,r}\|_F^2
	} .
\end{equation}

For two snapshots \(i\) and \(j\), we compute the distributed CSI similarity
\begin{equation}
	\label{eq:dcsi_similarity}
	s_{ij}
	=
	\frac{1}{R}
	\sum_{r=1}^{R}
	\frac{
	\left|
	\mathrm{tr}
	\left(
	\mathbf{C}_{i,r}^{\mathrm H}
	\mathbf{C}_{j,r}
	\right)
	\right|
	}{
	\|\mathbf{C}_{i,r}\|_F
	\|\mathbf{C}_{j,r}\|_F
	},
	\qquad
	d_{ij}=1-s_{ij}.
\end{equation}

For each training trajectory independently, we precompute the \(k_{+}=10\) nearest CSI neighbors of each snapshot \(i\) under \(d_{ij}\); these form its positive (teacher-neighbor) set \(\mathcal{P}(i)\). Restricting neighbors to the same trajectory avoids constructing positives across disconnected trajectories.

For a trajectory window \(\mathcal{W}\) in a mini-batch, we restrict \(\mathcal{P}(i)\) to the window:
\begin{equation}
\label{eq:window_positives}
    \mathcal{P}_{\mathcal{W}}(i)
	=
    \mathcal{P}(i)\cap\mathcal{W},
    \quad
    \mathcal{W}_{+}
	=
	\left\{
    i \in \mathcal{W} : \mathcal{P}_{\mathcal{W}}(i) \neq \varnothing
	\right\}.
\end{equation}

For each window with \(\mathcal{W}_{+}\neq\varnothing\), we use the weighted multi-positive contrastive loss
\begin{equation}
\label{eq:cc_final_loss}
    \mathcal{L}_{\mathrm{cc}}(\mathcal{W})
    =
    -\frac{1}{|\mathcal{W}_{+}|}
    \sum_{i \in \mathcal{W}_{+}}
    \log
    \frac{\sum_{j \in \mathcal{P}_{\mathcal{W}}(i)} w_{ij}\,e^{\mathrm{sim}(\mathbf{y}_c^{(i)},\mathbf{y}_c^{(j)})/\tau}}
         {\sum_{\ell \in \mathcal{W}\setminus\{i\}} e^{\mathrm{sim}(\mathbf{y}_c^{(i)},\mathbf{y}_c^{(\ell)})/\tau}} .
\end{equation}

where \(\mathrm{sim}(\cdot,\cdot)\) denotes cosine similarity, \(\tau=0.2\), and
\begin{equation}
\label{eq:cc_weights}
w_{ij}
=
\frac{\exp(-d_{ij}/\sigma_{\mathrm{cc}})}
{\sum_{m\in\mathcal{P}_{\mathcal{W}}(i)}\exp(-d_{im}/\sigma_{\mathrm{cc}})},
\qquad
\sigma_{\mathrm{cc}}=0.1 .
\end{equation}

The batch loss \(\mathcal{L}_{\mathrm{cc}}\) is the mean over all valid anchor terms from all valid windows. Windows with \(\mathcal{W}_{+}=\varnothing\) contribute zero.

The scalar \(\lambda_{\mathrm{cc}}\geq 0\) controls the strength of the channel-chart regularization. The dense full-report model is trained with
$\mathcal{L}_{\mathrm{dense}} = \mathcal{L}_{\mathrm{loc}} + \lambda_{\mathrm{cc}}\mathcal{L}_{\mathrm{cc}}$, where the localization term is evaluated on labeled samples and the charting term on both labeled and unlabeled samples.

For a mini-batch, the sparse objective is
\begin{equation}
    \label{eq:sparse_loss}
    \begin{aligned}
        \mathcal{L}_{\mathrm{sparse}}
        =
        &\;
         \mathcal{L}_{\mathrm{loc}}
        + 
        \lambda_{\mathrm{cc}}\mathcal{L}_{\mathrm{cc}}
        +    
        \lambda_{\mathrm{pool}}\mathcal{L}_{\mathrm{pool}}^{\mathrm{T}} \\
        &+
        \lambda_{\mathrm{chart}}\mathcal{L}_{\mathrm{chart}}^{\mathrm{T}}
        +
        \lambda_{\mathrm{pos}}\mathcal{L}_{\mathrm{pos}}^{\mathrm{T}}
        \\
        &+
        \lambda_{\mathrm{band}}\mathcal{L}_{\mathrm{band}}
        +    
        \lambda_{\mathrm{null}}\mathcal{L}_{\mathrm{null}}.
    \end{aligned}
\end{equation}

Here, \(\mathcal{L}_{\mathrm{loc}}\) and \(\mathcal{L}_{\mathrm{cc}}\) are the supervised localization and channel-chart losses already used for the dense model. The teacher terms \(\mathcal{L}_{\mathrm{pool}}^{\mathrm{T}}\), \(\mathcal{L}_{\mathrm{chart}}^{\mathrm{T}}\), and \(\mathcal{L}_{\mathrm{pos}}^{\mathrm{T}}\) distill the student's pooled representation, chart embedding, and position output toward the frozen dense teacher. The activity terms \(\mathcal{L}_{\mathrm{band}}\) and \(\mathcal{L}_{\mathrm{null}}\) keep the expected activity near the target budget and discourage samples in which all RAAs stay silent. The scalars \(\lambda_{\bullet}\ge 0\) weight these terms, with values in Section~\ref{sec:exp_protocol}; the chart terms are active only when charting is enabled (\(\lambda_{\mathrm{cc}}>0\)).

Each distillation term is a normalized squared error between a student quantity \(\mathbf{q}^{(n)}\) and its teacher counterpart \(\mathbf{q}^{T,(n)}\) (the superscript \(T\) denotes dense-teacher quantities),
\begin{equation}
    \label{eq:distill_generic}
    \mathcal{L}_{\mathrm{distill}}(\mathbf{q})
    =
    \frac{1}{|\mathcal{V}|}
    \sum_{n\in\mathcal{V}}
    \frac{1}{\dim(\mathbf{q})}
    \|\mathbf{q}^{(n)}-\mathbf{q}^{T,(n)}\|_2^2 ,
\end{equation}
applied to the pooled representation \(\mathbf{h}\) and chart embedding \(\mathbf{y}_c\) over all valid timesteps \(\mathcal{V}=\mathcal{V}_{\mathcal B}\), and to the localization output \(\hat{\mathbf{y}}\) over labeled timesteps \(\mathcal{V}=\mathcal{V}_{\mathcal B,L}\), giving \(\mathcal{L}_{\mathrm{pool}}^{\mathrm{T}}\), \(\mathcal{L}_{\mathrm{chart}}^{\mathrm{T}}\), and \(\mathcal{L}_{\mathrm{pos}}^{\mathrm{T}}\). We set \(\mathcal{L}_{\mathrm{pos}}^{\mathrm{T}}=0\) when \(|\mathcal{V}_{\mathcal B,L}|=0\).

For \(n\in\mathcal{V}_{\mathcal B}\), let \(m_\pi^{(n)}=\sum_{r=1}^{R}\pi_r^{(n)}\) denote the expected per-sample activity and \(\bar m_\pi\) its mean over valid timesteps. With band edges \(K_{\mathrm{lo}}=\max(0,K-\Delta_-)\) and \(K_{\mathrm{hi}}=K+\Delta_+\), the activity-band penalty keeps the expected activity within a soft interval around the target budget,
\begin{equation}
    \label{eq:band_loss}
    \mathcal{L}_{\mathrm{band}}
    =
    \frac{1}{|\mathcal{V}_{\mathcal B}|}
    \sum_{n\in\mathcal{V}_{\mathcal B}}
    \Bigl[
    \alpha_-\left(K_{\mathrm{lo}} - m_\pi^{(n)}\right)_+^2
    +
    \alpha_+\left(m_\pi^{(n)} - K_{\mathrm{hi}}\right)_+^2
    \Bigr] ,
\end{equation}
where \((x)_+=\max(x,0)\). The underuse weight \(\alpha_-\) is linearly annealed to zero over training while the overuse weight \(\alpha_+\) is fixed; for full-budget sparse runs an additional mean-activity cap \(\alpha_+(\bar m_\pi-\bar K_{\mathrm{hi}})_+^2\) is added to avoid collapse to dense reporting, where \(\bar K_{\mathrm{hi}}\) bounds the mean activity \(\bar m_\pi\). The null-report penalty
\begin{equation}
    \label{eq:null_loss}
    \mathcal{L}_{\mathrm{null}}
    =
    \frac{1}{|\mathcal{V}_{\mathcal B}|}
    \sum_{n\in\mathcal{V}_{\mathcal B}}
    \prod_{r=1}^{R}(1-\pi_r^{(n)})
\end{equation}
discourages samples for which all RAAs are likely to remain silent.


\section{Experimental Setup and Compared Methods}
\label{sec:exp_setup}

\subsection{Experimental Protocol}
\label{sec:exp_protocol}

Experiments are conducted on measured distributed CSI from industrial indoor scenarios in the DICHASUS dataset \cite{Euchner2023-wu}, with \(R=4\) RAAs, transmitted latent dimension \(d=64\), and auxiliary chart dimension \(T_c=16\). We use a trajectory-level sparse annotation split with separate DICHASUS recordings for labeled training, unlabeled training, validation and threshold calibration, and testing.\footnote{The recordings used are \texttt{cf02} (labeled training), \texttt{cf04} and \texttt{cf05} (unlabeled training), \texttt{cf07} (validation and threshold calibration), and \texttt{cf03} (test), all from the same measurement environment.} This yields \(N_L=18{,}602\) labeled and \(N_U=63{,}880\) unlabeled training snapshots (labeled fraction \(\approx 22.55\%\)), \(N_{\mathrm{val}}=59{,}345\) validation, and \(N_{\mathrm{test}}=22{,}643\) test snapshots. Test labels are used only for final evaluation. The trajectory-level partition prevents leakage across temporally adjacent samples and preserves a clean sequential protocol for stateful reporting policies.

All learned sparse models are trained under the same data split, preprocessing pipeline, and validation protocol. A separate sparse model is trained for each target activity budget \(K \in \{1, \ldots, R\}\). Sparse ETDI models are warm-started from a dense full-report teacher, trained with straight-through stochastic gates, and evaluated using validation-calibrated deterministic thresholds. The same validation trajectory is used to calibrate the decision thresholds of heuristic baselines. The activity-band penalty uses margins \((\Delta_-, \Delta_+) = (0.5, 0.25)\), with the underuse weight linearly annealed to zero during training while the overuse weight is fixed; the null-report term is weighted by \(\lambda_{\mathrm{null}} = 0.5\). Teacher distillation uses relative weights \(\lambda_{\mathrm{pool}} / \lambda_{\mathrm{chart}} / \lambda_{\mathrm{pos}} = 1 / 0.5 / 0.5\). Remaining settings, including epoch budgets, the anneal schedule, threshold-search grids, and the activity-cap overrides used at full budget, are documented in the code release \cite{narras_repo}. Reported comparisons are made at the realized test activity \(\bar M\), defined below.

We report localization accuracy alongside the realized communication cost, since \acro is explicitly activity-budgeted and the two trade off against each other. Localization accuracy is summarized by three statistics of the per-sample Euclidean error \(e^{(n)} = \|\hat{\mathbf{y}}^{(n)} - \mathbf{y}^{(n)}\|_2\): the mean Euclidean error \(\mathrm{MEDE} = \frac{1}{N_{\mathrm{test}}}\sum_{n} e^{(n)}\), the median error \(R50\), and the 90th-percentile error \(R90\). MEDE captures average performance, R50 typical-case behavior, and R90 the high-error tail (lower is better for all three). Realized communication cost is summarized by the per-sample number of active RAAs \(M^{(n)} = \sum_{r=1}^{R} c_r^{(n)}\), reported through its empirical mean \(\bar{M}\) and standard deviation \(s_M\). Trustworthiness and continuity are used as geometry diagnostics on held-out labeled data \cite{Studer2018-dk}.

\subsection{Compared Methods and Operating Points}
\label{sec:compared_methods}

The primary method evaluated in this paper is \acro, whose Recurrent Novelty Trigger (RNT) is introduced in Sec.~\ref{sec:proposed_solution}. To isolate the contribution of the proposed reporting policy, we compare against three groups of methods: (i) simplified ETDI trigger variants used for ablation, (ii) dense full-report references, and (iii) non-learned or projection-based sparse-reporting baselines. Whenever possible, methods share the same latent dimension \(d\), the same permutation-invariant fusion backbone, and the same validation protocol. All budgeted methods are evaluated across the same set of target budgets \(K \in \{1, \ldots, R\}\), enabling a direct comparison of accuracy--communication trade-offs.

\paragraph{Ablation variants}
We consider three variants that share the same encoder, fusion module, and prediction heads as \acro, isolating specific design choices.

The first variant is the \emph{Stateless Trigger} (ST), a memoryless policy that depends only on the current latent representation:
\begin{equation}
\label{eq:st_trigger}
    \pi_r^{(n)}
    =
    \sigma\!\left(
    \mathrm{MLP}_{\mathrm{stat}}
    \left(\mathbf{z}_r^{(n)}\right)
    \right),
\end{equation}
where \(\sigma(\cdot)\) is the logistic sigmoid. This tests whether instantaneous features alone suffice for the report/silence decision.

The \emph{EMA-Drift Trigger} (EDT) is a lightweight stateful policy that tracks a running exponential moving average (EMA) of recent latents,
\begin{equation}
\label{eq:ema_state}
    \mathbf{q}_r^{(n)}
    =
    \alpha \mathbf{q}_r^{(n-1)}
    +
    (1-\alpha)\mathbf{z}_r^{(n)},
    \quad 0 < \alpha < 1,
\end{equation}
with \(\mathbf{q}_r^{(0)} = \mathbf{0}\), and reports based on the drift from that reference:
\begin{equation}
\label{eq:ema_prob}
    \pi_r^{(n)}
    =
    \sigma\!\left(
    \mathrm{MLP}_{\mathrm{ema}}
    \left(
    \left\|
    \mathbf{z}_r^{(n)} -
    \mathbf{q}_r^{(n-1)}
    \right\|_2
    \right)
    \right).
\end{equation}
This captures local temporal change while remaining lighter than the full RNT.

Finally, the \emph{Localization-Only ETDI} (LO) is the full recurrent \acro architecture trained with \(\lambda_{\mathrm{cc}}=0\), removing charting regularization and chart distillation while retaining the same encoder--trigger--fusion architecture, dense warm start, and budgeted deployment calibration. The \(\lambda_{\mathrm{cc}}=0\) runs use the activity-selection overrides documented in the code release \cite{narras_repo}. This isolates the contribution of chart-aware training to localization accuracy and geometry preservation.

\paragraph{Dense and sparse baselines}

We compare against dense full-report references and several sparse baselines that do not learn a reporting policy but instead apply a fixed or simple decision rule. Each sparse baseline still compresses CSI into a per-RAA latent and reports it over the same fusion interface as \acro, so the methods differ only in the reporting rule (and, for LP+DT, in using a fixed PCA encoder); each reaches its target budget by calibrating a decision threshold on validation as detailed below, while URR meets the budget by construction. The baselines are:

\begin{itemize}[leftmargin=*]
    \item \emph{Dense Reporting Reference} (DRR): all RAAs report for every sample using a dense encoder--fusion model. We report three dense references: a chart-regularized teacher, a localization-only teacher, and a PCA-based backbone.
    \item \emph{Uniform Random Reporting} (URR): each RAA reports independently with probability $K/R$, without data-dependent decisions.
    \item \emph{Received-Power Trigger} (RPT): each RAA reports when a scalar received-power score exceeds a validation-calibrated threshold.
    \item \emph{Latent-Drift Trigger} (LDT): the dense non-charting backbone is reused, but reporting is determined by thresholding the drift $\|\mathbf{z}_r^{(n)} - \mathbf{j}_r^{(n-1)}\|_2$ relative to the last reported latent.
    \item \emph{Linear Projection + Drift Trigger} (LP+DT): the learned convolutional neural network (CNN) encoder is replaced by a fixed PCA projection fitted on preprocessed training CSI samples. Reporting then follows the same drift-based rule as LDT, using the projected latents.
\end{itemize}

\paragraph{Threshold calibration}
The sparse triggers ST, EDT, and \acro's RNT all output report probabilities \(\pi_r^{(n)}\), trained with a straight-through Bernoulli estimator. At validation and test time, the stochastic gate is replaced by $c_r^{(n)} = \mathbb{I} \{\pi_r^{(n)}\geq \eta_K\}$,

where \(\eta_K\) is recalibrated on the validation trajectory under a budget-first rule that prefers candidates satisfying the activity and null-report constraints and otherwise falls back to a penalized validation objective. The deterministic baselines RPT, LDT, and LP+DT threshold a scalar score (received power, latent drift, or PCA-projected drift) using the same budget-first principle. URR samples reports with probability \(K/R\); DRR has no threshold. Full grid sizes, checkpoint-selection criteria, and the chart-regularized geometry term used during selection are documented in the public code repository \cite{narras_repo}.


\section{Results}
\label{sec:results}

This section evaluates the proposed ETDI framework under the activity-budgeted setting defined in Sections~\ref{sec:sys_model}--\ref{sec:exp_setup}. The main claim tested here is that the learned decentralized reporting employed by \acro provides a better localization-versus-activity trade-off than competing sparse-reporting policies at comparable realized activity. All reported results use the same seed and trajectory split described in Section~\ref{sec:exp_protocol}.

\subsection{Communication--accuracy Trade-off}
\label{sec:comm_accuracy_results}

\begin{figure}[t]
	\centering
	\includegraphics[width=\columnwidth]{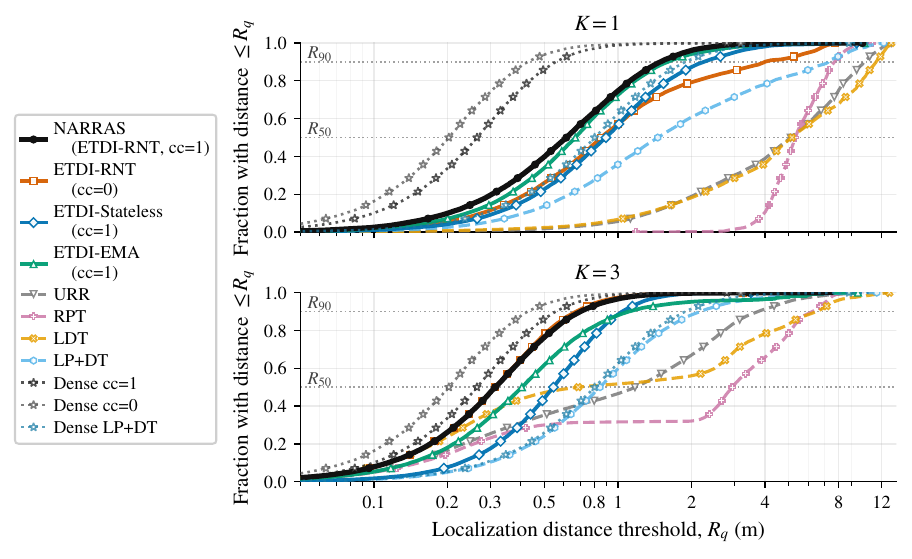}
    \vspace{-0.8cm}
     	\caption{Empirical CDFs of the localization error for representative activity budgets \(K=1\) and \(K=3\). For each distance threshold \(R_q\), the vertical axis reports the fraction of test samples with localization error no larger than \(R_q\). Horizontal dotted lines mark the median and 90th-percentile levels, corresponding to \(R_{50}\) and \(R_{90}\). Left-shifted and steeper curves indicate lower typical error and a smaller high-error tail. Dense references are shown as budget-free full-report baselines.
	}
	\label{fig:error_cdf_k1_k3}
 \end{figure}

The primary comparison concerns the trade-off between localization accuracy and reporting activity. Tables~\ref{tab:main_budget_table}--\ref{tab:sparse_baselines} summarize this trade-off through the empirical average activity \(\bar{M}\), the mean Euclidean distance error (MEDE), and the median and 90th-percentile localization errors. To complement these aggregate statistics, Fig.~\ref{fig:error_cdf_k1_k3} reports the empirical cumulative distribution functions (CDFs) of the localization error for representative low- and moderate-activity budgets (\(K=1\) and \(K=3\)). This view is useful because two methods with similar mean error may differ substantially in their typical error or high-error tail.

Throughout the result tables, \(cc=1\) and \(cc=0\) denote \(\lambda_{\mathrm{cc}}=1\) and \(\lambda_{\mathrm{cc}}=0\), respectively.

Table~\ref{tab:main_budget_table} reports the chart-regularized ETDI variants. The recurrent novelty trigger is most beneficial in the low- and moderate-activity regimes. At \(K=1\), ETDI--RNT obtains a mean error of \(0.7675\) and \(R90\) of \(1.4856\), markedly better than both the stateless and EMA triggers. The CDFs in Fig.~\ref{fig:error_cdf_k1_k3} show that this improvement is not only a mean-error effect: the \acro curve is shifted to the left over most of the distribution and reaches both the \(R50\) and \(R90\) levels at smaller distance thresholds. At \(K=2\) and \(K=3\), the recurrent trigger remains among the strongest sparse policies, while at near-full activity (\(K=4\)) the differences shrink because almost all RAAs are active.

\begin{table}[t]
    \centering
    \caption{
        Main chart-regularized {ETDI} comparison. All models use the same
        shared encoder--fusion architecture and the dense cc\,=\,1 teacher;
        only the decentralized reporting trigger changes.
    }
    \label{tab:main_budget_table}
    \begin{tabular}{@{}lcccccc@{}}
        \toprule
        Method & $K$ & $\bar{M}$ & $s_M$ & MEDE & R50 & R90 \\
        \midrule
        \multicolumn{7}{@{}l@{}}{\emph{Recurrent Novelty Trigger (RNT)}} \\[1pt]
        \quad cc\,=\,1 & 1 & 1.0136 & 0.1715 & \textbf{0.7675} & \textbf{0.6137} & \textbf{1.4856} \\
        \quad cc\,=\,1 & 2 & 1.9725 & 0.3184 & \textbf{0.4675} & \textbf{0.3773} & \textbf{0.8295} \\
        \quad cc\,=\,1 & 3 & 2.9320 & 0.2755 & \textbf{0.3796} & \textbf{0.3138} & \textbf{0.7013} \\
        \quad cc\,=\,1 & 4 & 3.8710 & 0.4314 & 0.3431 & \textbf{0.2746} & 0.6273 \\
        \midrule
        \multicolumn{7}{@{}l@{}}{\emph{Stateless Trigger (ST)}} \\[1pt]
        \quad cc\,=\,1 & 1 & 1.1256 & 0.5422 & 1.1410 & 0.8919 & 2.2371 \\
        \quad cc\,=\,1 & 2 & 2.0531 & 0.4889 & 0.6882 & 0.6165 & 1.2081 \\
        \quad cc\,=\,1 & 3 & 2.9575 & 0.6082 & 0.6025 & 0.5436 & 1.0400 \\
        \quad cc\,=\,1 & 4 & 3.9710 & 0.1721 & \textbf{0.3308} & 0.2866 & \textbf{0.5975} \\
        \midrule
        \multicolumn{7}{@{}l@{}}{\emph{EMA-Drift Trigger (EDT)}} \\[1pt]
        \quad cc\,=\,1 & 1 & 1.2543 & 0.6434 & 0.8513 & 0.6688 & 1.5982 \\
        \quad cc\,=\,1 & 2 & 2.2551 & 1.0438 & 0.6985 & 0.5101 & 1.2959 \\
        \quad cc\,=\,1 & 3 & 3.2002 & 1.0167 & 0.6801 & 0.4037 & 1.0555 \\
        \quad cc\,=\,1 & 4 & 3.8935 & 0.3144 & 0.3582 & 0.3125 & 0.6330 \\
        \bottomrule
    \end{tabular}
\end{table}

Table~\ref{tab:dense_teachers} reports the full-report references used as budget-free upper bounds: the chart-regularized dense teacher, the localization-only dense teacher, and the PCA-based dense backbone used by LP+DT.

\begin{table}[t]
	\centering
	\caption{Dense full-report references. All RAAs are active for every sample.}		
	\label{tab:dense_teachers}
	\footnotesize
	\setlength{\tabcolsep}{4pt}
	\begin{tabular}{lccc}
		\toprule
		Dense reference & MEDE & R50 & R90 \\
		\midrule
		Dense cc=1 & 0.3080 & 0.2642 & 0.5567 \\
		Dense cc=0 & \textbf{0.2396} & \textbf{0.2027} & \textbf{0.4271} \\
		Dense LP+DT & 0.9904 & 0.8012 & 1.8840 \\
		\bottomrule
	\end{tabular}
\end{table}

\subsection{Effect of Chart Regularization}
\label{sec:chart_ablation_results}

Table~\ref{tab:rnt_cc_ablation} compares the recurrent trigger with and without chart regularization. The chart-regularized RNT is substantially more robust in the low-budget regime: at \(K=1\), enabling charting reduces MEDE from \(1.5140\) to \(0.7675\) and \(R90\) from \(3.9808\) to \(1.4856\), with a similar improvement at \(K=2\). At \(K=3\) and \(K=4\), the no-chart model is slightly lower in mean error, but it is initialized from the localization-only dense teacher, which is itself more accurate than the chart-regularized dense teacher, and \(K=4\) is close to the dense regime. We therefore interpret the cc=0 comparison as an ablation, not as the main operating mode.

\begin{table}[t]
	\centering
	\caption{
		Recurrent-trigger ablation with and without channel-chart regularization. The cc=1 model uses the chart-regularized dense teacher, while the cc=0 model uses the localization-only dense teacher.
	}
	\label{tab:rnt_cc_ablation}
	\footnotesize
	\setlength{\tabcolsep}{4pt}
    \begin{tabular}{lcccccc}
		\toprule
		Method & \(K\) & \(\bar{M}\) & \(s_M\) & MEDE & R50 & R90 \\
		 \midrule
		ETDI--RNT, cc=1 & 1 & 1.0136 & 0.1715 & \textbf{0.7675} & \textbf{0.6137} & \textbf{1.4856} \\
		ETDI--RNT, cc=1 & 2 & 1.9725 & 0.3184 & \textbf{0.4675} & \textbf{0.3773} & \textbf{0.8295} \\
		ETDI--RNT, cc=1 & 3 & 2.9320 & 0.2755 & 0.3796 & \textbf{0.3138} & 0.7013 \\
		ETDI--RNT, cc=1 & 4 & 3.8710 & 0.4314 & 0.3431 & \textbf{0.2746} & 0.6273 \\
		\midrule
		ETDI--RNT, cc=0 & 1 & 1.0730 & 0.8182 & 1.5140 & 0.8486 & 3.9808 \\
		ETDI--RNT, cc=0 & 2 & 1.9721 & 0.8658 & 0.6127 & 0.4142 & 1.2910 \\
		ETDI--RNT, cc=0 & 3 & 2.9371 & 0.7759 & \textbf{0.3683} & 0.3140 & \textbf{0.6733} \\
		ETDI--RNT, cc=0 & 4 & 3.7858 & 0.4429 & \textbf{0.3271} & 0.2880 & \textbf{0.5782} \\
		\bottomrule
	\end{tabular}
\end{table}

The comparison should be read as a chart-regularization ablation rather than a strict same-teacher comparison: the cc=0 model is initialized from a stronger localization-only dense teacher (Table~\ref{tab:dense_teachers}), yet the chart-regularized sparse model is substantially better at \(K=1\) and \(K=2\), especially in R90. Geometry-aware regularization is therefore most valuable precisely when the reporting budget is tight and the risk of large errors is highest.

The chart objective also stabilizes the local channel geometry under sparse reporting. Evaluated as continuity and trustworthiness on held-out labeled samples against ground-truth positions, both metrics stay close to the dense chart-regularized reference and vary only mildly with the activity budget, indicating that sparse reporting does not erase the local neighborhood structure of the learned representation. These diagnostics are largely insensitive to the reporting policy; the localization gains in Tables~\ref{tab:main_budget_table}--\ref{tab:rnt_cc_ablation} remain the primary evidence for the benefit of the recurrent trigger.

\subsection{Comparison with Non-charted Sparse Baselines}
\label{sec:sparse_baseline_results}

Table~\ref{tab:sparse_baselines} reports the non-charted sparse baselines. For these methods, the charting objective is disabled and chart outputs are not interpreted, so charting metrics are not reported. Among these baselines, LP+DT is the strongest at low and moderate budgets, yet ETDI--RNT with chart regularization remains far more accurate: at \(K=1\), LP+DT obtains MEDE/\(R90\) of \(2.6069/7.2528\), against \(0.7675/1.4856\) for ETDI--RNT cc=1. The CDFs in Fig.~\ref{fig:error_cdf_k1_k3} make the same effect visible distributionally: the non-learned and projection-based sparse baselines are markedly right-shifted, especially at \(K=1\), indicating both larger typical errors and substantially heavier tails.

\begin{table}[t]
	\centering
	\caption{
		Non-charted sparse-reporting baselines. For these methods, the charting objective is disabled and charting metrics are not reported. URR at \(K=4\) is omitted as it corresponds to the dense case.
	}
	\label{tab:sparse_baselines}
	\footnotesize
	\setlength{\tabcolsep}{4pt}
	\begin{tabular}{lcccccc}
		\toprule
		Method & \(K\) & \(\bar{M}\) & \(s_M\) & MEDE & R50 & R90 \\
		\midrule
		URR & 1 & 1.0018 & 0.8680 & 5.4533 & 5.0947 & 10.2220 \\
		URR & 2 & 1.9944 & 0.9984 & 3.2217 & 2.4444 & 7.1778 \\
		URR & 3 & 2.9960 & 0.8659 & 1.6262 & 1.1723 & 3.7269 \\
		\midrule
		RPT & 1 & 1.1186 & 0.8231 & 5.6179 & 5.3188 & 7.6761 \\
		RPT & 2 & 1.9657 & 0.6454 & 4.1407 & 4.3250 & 6.4468 \\
		RPT & 3 & 2.9160 & 0.8420 & 2.9901 & 2.9320 & 6.2710 \\
		RPT & 4 & 3.4897 & 0.6455 & 1.5445 & 0.3825 & 3.6190 \\
		\midrule
		LDT & 1 & 1.0383 & 0.9925 & 5.8788 & 5.1542 & 11.3901 \\
		LDT & 2 & 2.1450 & 1.3394 & 3.9966 & 3.2845 & 9.5956 \\
		LDT & 3 & 3.0798 & 0.4892 & 2.2857 & 0.6885 & 6.4783 \\
		LDT & 4 & 3.7445 & 0.5847 & \textbf{0.7893} & \textbf{0.2400} & 2.7221 \\
		\midrule
		LP+DT & 1 & 1.1967 & 0.8131 & \textbf{2.6069} & \textbf{1.4526} & \textbf{7.2528} \\
		LP+DT & 2 & 2.0192 & 0.9540 & \textbf{1.6565} & \textbf{0.9912} & \textbf{3.5571} \\
		LP+DT & 3 & 3.0470 & 0.9615 & \textbf{1.1294} & \textbf{0.8430} & \textbf{2.1640} \\
		LP+DT & 4 & 3.7069 & 0.6167 & 1.0094 & 0.8102 & \textbf{1.9208} \\
		\bottomrule
	\end{tabular}
\end{table}

\subsection{Deployment Overhead}
\label{sec:overhead_results}

The proposed architecture remains lightweight enough for the considered IoT setting. Each RAA runs the shared convolutional encoder (\(221{,}216\) parameters) and, for \acro, the recurrent trigger adds only \(29{,}249\) parameters and \(128\) state variables per RAA, holding the recurrent state and the last reported latent. Each active report carries a \(64\)-dimensional latent vector, and the fusion center uses \(72{,}915\) parameters for RAA embeddings, attention-based set aggregation, the empty-set representation, and the prediction heads. The teacher networks are used only during training. This small trigger footprint makes \acro practical for resource-constrained edge nodes. Moreover, since radio transmission dominates energy consumption in many low-power devices, reducing the average number of transmitted latent reports provides a first-order proxy for energy expenditure. These quantities are model-size and latent-payload indicators; a bit-level communication analysis is outside the scope of this work.

\subsection{Spatial Distribution of Localization Error}
\label{sec:spatial_error_results}

Aggregate error statistics do not reveal where localization errors occur in the physical environment. We therefore analyze the spatial distribution of the error for the proposed chart-regularized recurrent policy. Let \(\mathcal{C}\) denote a spatial grid cell and let \(\mathcal{I}(\mathcal{C}) = \{n : \mathbf{y}^{(n)} \in \mathcal{C}\}\) be the set of labeled test samples whose ground-truth position lies in that cell.

The cell-wise mean localization error is
 \begin{equation}
	E(\mathcal{C})
	=
	\frac{1}{|\mathcal{I}(\mathcal{C})|}
	\sum_{n \in \mathcal{I}(\mathcal{C})}
	\left\|
	\hat{\mathbf{y}}^{(n)}
	-
	\mathbf{y}^{(n)}
	\right\|_2,
 \end{equation}

for cells with \(|\mathcal{I}(\mathcal{C})|>0\).

\begin{figure}[t]
     \centering
	\includegraphics[width=\columnwidth]{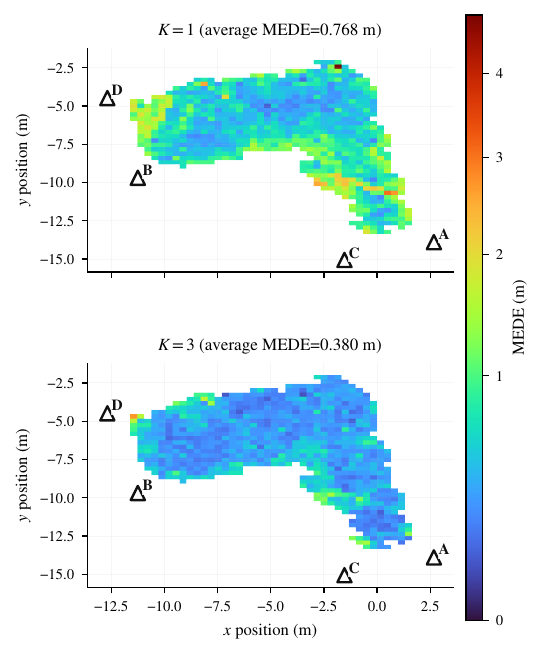}
    \vspace{-1cm}
     \caption{
	Spatial distribution of the localization error for \acro under target budgets \(K=1\) and \(K=3\). Each occupied grid cell is colored by \(E(\mathcal{C})\), the mean Euclidean localization error of the test samples whose ground-truth positions fall in that cell. Triangles mark the RAA locations, and the panel titles report the corresponding test-set MEDE. Increasing the budget from \(K=1\) to \(K=3\) reduces the error over most of the measured region and attenuates localized high-error pockets.
	}
	\label{fig:spatial_error_k1_k3}
 \end{figure}

Figure~\ref{fig:spatial_error_k1_k3} reports \(E(\mathcal{C})\) for \acro at \(K=1\) and \(K=3\). Increasing the activity budget reduces the error over most of the occupied test region, showing that the gain from \(K=1\) to \(K=3\) is spatially widespread rather than concentrated in a small subset of samples. The remaining high-error pockets appear near boundary and transition regions of the measured trajectory, where changes in the active RAA subset may have a larger effect on the fused representation, consistent with the tail behavior seen in the CDFs of Fig.~\ref{fig:error_cdf_k1_k3}.

\section{Conclusion}
\label{sec:conclusion}

This paper introduced \acro, a dense-to-sparse ETDI framework for CSI-based localization with decentralized reporting over a shared uplink. Each RAA makes a local reporting decision from its current latent representation, a recurrent local state, and a memory of the last transmitted latent, while the fusion center aggregates the resulting variable-cardinality set of reports through a permutation-invariant model. The training pipeline couples supervised localization, channel-chart regularization, dense-teacher distillation, and validation-calibrated activity control within a single learning objective.

Experiments on measured DICHASUS distributed CSI show that the recurrent novelty trigger is most effective in the communication-limited regime, where learning selective RAA reporting is essential. The chart-regularized variant is substantially more robust at the tightest budgets, with a clear improvement in the high-error tail, while geometry diagnostics remain stable across activity budgets. The CDF and spatial-error analyses indicate that these gains are not confined to average performance: the proposed policy shifts the error distribution toward smaller errors and attenuates localized high-error regions as the budget increases. At near-full activity the differences across triggers shrink, as expected, since the problem approaches dense reporting.

The present study focuses on the learning and inference aspects of decentralized reporting and abstracts the shared uplink through an average activity budget. The reported payloads are latent real-valued vectors rather than quantized packets, and the model does not explicitly simulate collisions, modulation, scheduling, latency, or energy consumption. Reduced reporting activity can nonetheless decrease contention over shared grant-free access channels, potentially lowering access latency in delay-sensitive vehicular deployments. Natural next steps are to integrate \acro with physical-layer random-access models, bit-level quantization, and packet-level collision handling, and to evaluate robustness under broader mobility patterns and heterogeneous edge-device constraints.

\section*{Code Availability}
\label{sec:code_availability}
A reference implementation of \acro, including the dense-to-sparse training pipeline, hyperparameter configurations, threshold-calibration procedure, and evaluation scripts, is publicly available at \cite{narras_repo}.







\bibliographystyle{IEEEtran}
\bibliography{references}

@ARTICLE{Guo2022-ih,
  author  = {Guo, Jiajia and Wen, Chao-Kai and Jin, Shi and Li, Geoffrey Ye},
  title   = {Overview of deep learning-based {CSI} feedback in massive {MIMO} systems},
  journal = {IEEE Trans. Commun.},
  volume  = {70},
  number  = {12},
  pages   = {8017--8045},
  year    = {2022},
  month   = dec,
  doi     = {10.1109/tcomm.2022.3217777},
}

@INPROCEEDINGS{Miretti2024-ge,
  author    = {Miretti, Lorenzo and Stańczak, Sławomir},
  title     = {Unlocking the potential of local {CSI} in cell-free networks with channel aging and fronthaul delays},
  booktitle = {Proc. {IEEE} Int. Conf. Commun. ({ICC})},
  pages     = {3761--3766},
  year      = {2024},
  month     = jun,
  doi       = {10.1109/icc51166.2024.10622732},
}

@ARTICLE{Djosic2021-hl,
  author  = {Djosic, Sandra and Stojanovic, Igor and Jovanovic, Milica and Nikolic, Tatjana and Djordjevic, Goran Lj},
  title   = {Fingerprinting-assisted {UWB}-based localization technique for complex indoor environments},
  journal = {Expert Syst. Appl.},
  volume  = {167},
  pages   = {114188},
  year    = {2021},
  month   = apr,
  doi     = {10.1016/j.eswa.2020.114188},
}

@ARTICLE{Mobini2026-uo,
  author  = {Mobini, Zahra and Gokceoglu, Ahmet Hasim and Wang, Li and Peters, Gunnar and Shin, Hyundong and Ngo, Hien Quoc},
  title   = {Cluster-wise processing in fronthaul-aware cell-free massive {MIMO} systems},
  journal = {IEEE Trans. Wireless Commun.},
  volume  = {25},
  pages   = {6240--6254},
  year    = {2026},
  doi     = {10.1109/twc.2025.3624199},
}

@ARTICLE{Taner2025-jo,
  author={Taner, Sueda and Palhares, Victoria and Studer, Christoph},
  journal={IEEE Trans. Wireless Commun.}, 
  title={Channel Charting in Real-World Coordinates With Distributed {MIMO}}, 
  year={2025},
  month={Apr.},
  volume={24},
  number={9},
  pages={7286-7300},
  doi={10.1109/TWC.2025.3559723}}

@INPROCEEDINGS{Lei2019-qr,
  author    = {Lei, Eric and Castañeda, Oscar and Tirkkonen, Olav and Goldstein, Tom and Studer, Christoph},
  title     = {Siamese neural networks for wireless positioning and channel charting},
  booktitle = {Proc. Annu. Allerton Conf. Commun. Control Comput.},
  pages     = {200--207},
  year      = {2019},
  month     = sep,
  doi       = {10.1109/allerton.2019.8919897},
}

@ARTICLE{Foliadis2024-yn,
  author  = {Foliadis, Anastasios and Castañeda Garcia, Mario H and Stirling-Gallacher, Richard A and Thomä, Reiner S},
  title   = {Deep learning-based positioning with multi-task learning and uncertainty-based fusion},
  journal = {IEEE Trans. Mach. Learn. Commun. Netw.},
  volume  = {2},
  pages   = {1127--1141},
  year    = {2024},
  doi     = {10.1109/tmlcn.2024.3441521},
}

@ARTICLE{Garcia-Saavedra2018-fn,
  author  = {Garcia-Saavedra, Andres and Salvat, Josep Xavier and Li, Xi and Costa-Perez, Xavier},
  title   = {{WizHaul}: on the centralization degree of cloud {RAN} next generation fronthaul},
  journal = {IEEE Trans. Mobile Comput.},
  volume  = {17},
  number  = {10},
  pages   = {2452--2466},
  year    = {2018},
  month   = oct,
  doi     = {10.1109/tmc.2018.2793859},
}

@INPROCEEDINGS{Wiffen2021-cn,
  author    = {Wiffen, Fred and Chin, Woon Hau and Doufexi, Angela},
  title     = {Distributed dimension reduction for distributed massive {MIMO} {C}-{RAN} with finite fronthaul capacity},
  booktitle = {Proc. Asilomar Conf. Signals Syst. Comput.},
  pages     = {1228--1236},
  year      = {2021},
  month     = oct,
  doi       = {10.1109/ieeeconf53345.2021.9723180},
}

@ARTICLE{Wu2013-zi,
  author  = {Wu, Kaishun and Xiao, Jiang and Yi, Youwen and Chen, Dihu and Luo, Xiaonan and Ni, Lionel M},
  title   = {{CSI}-based indoor localization},
  journal = {IEEE Trans. Parallel Distrib. Syst.},
  volume  = {24},
  number  = {7},
  pages   = {1300--1309},
  year    = {2013},
  month   = jul,
  doi     = {10.1109/tpds.2012.214},
}

@ARTICLE{Larsson2014-bf,
  author  = {Larsson, Erik G and Edfors, Ove and Tufvesson, Fredrik and Marzetta, Thomas L},
  title   = {Massive {MIMO} for next generation wireless systems},
  journal = {IEEE Commun. Mag.},
  volume  = {52},
  number  = {2},
  pages   = {186--195},
  year    = {2014},
  month   = feb,
  doi     = {10.1109/mcom.2014.6736761},
}

@misc{Che2024-ma,
  author        = {Che, Haohong and You, Li and Wang, Jue and Jin, Zhenzhou and Xie, Chenjie and Gao, Xiqi},
  title         = {Channel charting-assisted non-orthogonal pilot allocation for uplink {XL}-{MIMO} transmission},
  year          = {2024},
  month         = dec,
  eprint        = {2412.20920},
  archivePrefix = {arXiv},
  primaryClass  = {cs.IT},
}

@misc{Palhares2025-wg,
  author        = {Palhares, Victoria and Taner, Sueda and Studer, Christoph},
  title         = {{CSI2Vec}: towards a universal {CSI} feature representation for positioning and channel charting},
  year          = {2025},
  month         = jun,
  eprint        = {2506.05237},
  archivePrefix = {arXiv},
  primaryClass  = {cs.IT},
}

@ARTICLE{Stephan2024-tw,
  author  = {Stephan, Phillip and Euchner, Florian and Brink, Stephan Ten},
  title   = {Angle-delay profile-based and timestamp-aided dissimilarity metrics for channel charting},
  journal = {IEEE Trans. Commun.},
  volume  = {72},
  number  = {9},
  pages   = {5611--5625},
  year    = {2024},
  month   = sep,
  doi     = {10.1109/tcomm.2024.3386571},
}

@INPROCEEDINGS{Ribeiro2020-gy,
  author    = {Ribeiro, Lucas and Leinonen, Markus and Djelouat, Hamza and Juntti, Markku},
  title     = {Channel charting for pilot reuse in {mMTC} with spatially correlated {MIMO} channels},
  booktitle = {Proc. {IEEE} Global Commun. Conf. Workshops ({GC Wkshps})},
  year      = {2020},
  month     = dec,
  doi       = {10.1109/gcwkshps50303.2020.9367434},
}

@MISC{Euchner2023-wu,
  author       = {Euchner, Florian and Lee, Chae Eun and Stephan, Phillip},
  title        = {Dissimilarity metric-based channel charting},
  year         = {2023},
  howpublished = {\url{https://dichasus.inue.uni-stuttgart.de/tutorials/tutorial/dissimilarity-metric-channelcharting/}},
  note         = {Accessed: 2025-11-18},
}

@INPROCEEDINGS{Euchner2021-mx,
  author    = {Euchner, Florian and Gauger, Marc and Doerner, Sebastian and ten Brink, Stephan},
  title     = {A distributed massive {MIMO} channel sounder for ``big {CSI} data''-driven machine learning},
  booktitle = {Proc. 25th Int. {ITG} Workshop Smart Antennas ({WSA})},
  year      = {2021},
  month     = nov,
}

@INPROCEEDINGS{Euchner2022-am,
  author    = {Euchner, Florian and Stephan, Phillip and Gauger, Marc and Dörner, Sebastian and Brink, Stephan Ten},
  title     = {Improving triplet-based channel charting on distributed massive {MIMO} measurements},
  booktitle = {Proc. {IEEE} Int. Workshop Signal Process. Adv. Wireless Commun. ({SPAWC})},
  year      = {2022},
  month     = jul,
  doi       = {10.1109/spawc51304.2022.9833925},
}

@INPROCEEDINGS{Salihu2022-hk,
  author    = {Salihu, Artan and Schwarz, Stefan and Rupp, Markus},
  title     = {Learning-based remote radio head selection and localization in distributed antenna system},
  booktitle = {Proc. Jt. {Eur.} Conf. Netw. Commun. 6G Summit ({EuCNC}/6G Summit)},
  pages     = {65--70},
  year      = {2022},
  month     = jun,
  doi       = {10.1109/eucnc/6gsummit54941.2022.9815773},
}

@ARTICLE{Liu2018-qh,
  author  = {Liu, Liang and Larsson, Erik G and Yu, Wei and Popovski, Petar and Stefanovic, Cedomir and de Carvalho, Elisabeth},
  title   = {Sparse signal processing for grant-free massive connectivity: a future paradigm for random access protocols in the internet of things},
  journal = {IEEE Signal Process. Mag.},
  volume  = {35},
  number  = {5},
  pages   = {88--99},
  year    = {2018},
  month   = sep,
  doi     = {10.1109/msp.2018.2844952},
}

@ARTICLE{Studer2018-dk,
  author  = {Studer, Christoph and Medjkouh, Said and Gonultas, Emre and Goldstein, Tom and Tirkkonen, Olav},
  title   = {Channel charting: locating users within the radio environment using channel state information},
  journal = {IEEE Access},
  volume  = {6},
  pages   = {47682--47698},
  year    = {2018},
  doi     = {10.1109/access.2018.2866979},
}

@INPROCEEDINGS{Ferrand2020-of,
  author    = {Ferrand, Paul and Decurninge, Alexis and Ordoñez, Luis G and Guillaud, Maxime},
  title     = {Triplet-based wireless channel charting},
  booktitle = {Proc. {IEEE} Global Commun. Conf. ({GLOBECOM})},
  year      = {2020},
  month     = dec,
  doi       = {10.1109/globecom42002.2020.9322643},
}

@INPROCEEDINGS{Medjkouh2018-sh,
  author    = {Medjkouh, Said and Gonultas, Emre and Goldstein, Tom and Tirkkonen, Olav and Studer, Christoph},
  title     = {Unsupervised charting of wireless channels},
  booktitle = {Proc. {IEEE} Global Commun. Conf. ({GLOBECOM})},
  year      = {2018},
  month     = dec,
  doi       = {10.1109/glocom.2018.8647535},
}

@misc{narras_repo,
  author       = {Oliver, R. and Vazquez Alvarez, R. and Lancho, A. and Rini, S.},
  title        = {{NARRAS}: Reference implementation},
  year         = {2026},
  howpublished = {\url{https://github.com/roliverc/narras-csi-localization}},
  note         = {Accessed: 2026-05-28}
}


 



\vfill

\end{document}